\documentclass[12pt]{article}
\usepackage{latexsym}
\usepackage{amsmath}
\usepackage{amssymb}
\usepackage{relsize}
\usepackage{geometry}
\def\beqn{\begin{eqnarray}}
\def\eeqn{\end{eqnarray}}

\def\beq{\begin{equation}}
\def\eeq{\end{equation}}
\def\ba{\beq\new\begin{array}{c}}
\def\ea{\end{array}\eeq}

\newcommand{\gsim}{\lower.7ex\hbox{$
\;\stackrel{\textstyle>}{\sim}\;$}}
\newcommand{\lsim}{\lower.7ex\hbox{$
\;\stackrel{\textstyle<}{\sim}\;$}}

\newcommand{\ntwo}{${\mathcal N}=2$ }

\newcommand{\ntwot}{${\mathcal N}= \left(2,2\right) $ }
\newcommand{\ntwoo}{${\mathcal N}= \left(0,2\right) $ }
\newcommand{\ntwoon}{${\mathcal N}= \left(0,2\right) $}
\newcommand{\none}{${\mathcal N}=1$ }
\newcommand{\nonen}{${\mathcal N}=1$}

\newcommand{\pt}{\partial}

\renewcommand{\theequation}{\thesection.\arabic{equation}}

\newcommand{\p}{\partial}
\newcommand{\wt}{\widetilde}
\newcommand{\ov}{\overline}
\newcommand{\mc}[1]{\mathcal{#1}}

\newcommand{\lgr}{\left\lgroup}
\newcommand{\rgr}{\right\rgroup}

\def\slashed#1{\setbox0=\hbox{$#1$}             
   \dimen0=\wd0                                 
   \setbox1=\hbox{/} \dimen1=\wd1               
   \ifdim\dimen0>\dimen1                        
      \rlap{\hbox to \dimen0{\hfil/\hfil}}      
      #1                                        
   \else                                        
      \rlap{\hbox to \dimen1{\hfil$#1$\hfil}}   
      /                                         
   \fi}                                        %

\newcommand{\bxir}{\ov{\xi}{}_R}
\newcommand{\bxil}{\ov{\xi}{}_L}
\newcommand{\xir}{\xi_R}
\newcommand{\xil}{\xi_L}

\newcommand{\bzr}{\ov{\zeta}{}_R}
\newcommand{\zr}{\zeta_R}

\newcommand{\nbar}{\ov{n}}

\newcommand{\ssm}{{\scriptscriptstyle(M)}}
\newcommand{\sse}{{\scriptscriptstyle(E)}}
\newcommand{\cell}{{\mathcal L}}
\newcommand{\CPC}{CP($N-1$)$\times$C }

\newcommand{\cpn}{CP$(N-1)\,$}

\newcommand{\lar}{\lambda_R}
\newcommand{\lal}{\lambda_L}
\newcommand{\larl}{\lambda_{R,L}}

\newcommand{\blar}{\ov{\lambda}{}_R}
\newcommand{\blal}{\ov{\lambda}{}_L}
\newcommand{\blarl}{\ov{\lambda}{}_{R,L}}

\newcommand{\tgamma}{\wt{\gamma}}
\newcommand{\btgamma}{\ov{\tgamma}}
\newcommand{\bpsi}{\ov{\psi}{}}
\newcommand{\bphi}{\ov{\phi}{}}
\newcommand{\bxi}{\ov{\xi}{}}

\newcommand{\ff}{\mc{F}}
\newcommand{\bff}{\ov{\mc{F}}}

\newcommand{\eer}{\epsilon_R}
\newcommand{\eel}{\epsilon_L}
\newcommand{\eerl}{\epsilon_{R,L}}

\newcommand{\beer}{\ov{\epsilon}{}_R}
\newcommand{\beel}{\ov{\epsilon}{}_L}
\newcommand{\beerl}{\ov{\epsilon}{}_{R,L}}

\newcommand{\bi}{{\bar \imath}}
\newcommand{\bj}{{\bar \jmath}}
\newcommand{\bk}{{\bar k}}
\newcommand{\bl}{{\bar l}}
\newcommand{\bm}{{\bar m}}

\begin{document}

\begin{titlepage}

\begin{flushright}
FTPI-MINN-09/22,  UMN-TH-2802/09\\
July 2, 2009
\end{flushright}

\vspace{2.0cm}

\begin{center}

{\Large \bf   Heterotic \boldmath{\ntwoon} \boldmath{${\rm CP}(N-1)$}
model \\ with twisted masses
 }

\end{center}

\begin{center}
{\bf P. A. Bolokhov$^{a,b}$, M.~Shifman$^{c,d}$ and \bf A.~Yung$^{c,d,e}$}
\end {center}
\vspace{0.3cm}
\begin{center}

$^a${\it Physics and Astronomy Department, University of Pittsburgh, Pittsburgh, Pennsylvania, 15260, USA}\\
$^b${\it Theoretical Physics Department, St.Petersburg State University, Ulyanovskaya~1, 
	 Peterhof, St.Petersburg, 198504, Russia}\\
$^c${\it  William I. Fine Theoretical Physics Institute,
University of Minnesota,
Minneapolis, MN 55455, USA}\\
$^d${\it Institut de Physique Th\'eorique, CEA Saclay, 91191 Gif-sur-Yvette C\'edex, France}\\
$^{e}${\it Petersburg Nuclear Physics Institute, Gatchina, St. Petersburg
188300, Russia}\\

\end{center}

\vspace{1cm}

\begin{abstract}

We present a two-dimensional heterotic \ntwoo CP$(N-1)$ model
with twisted masses. It is supposed to describe internal dynamics of non-Abelian strings 
in massive \ntwo SQCD with
\none-preserving deformations. We present gauge and geometric
formulations of the world-sheet theory and check its \ntwoo
supersymmetry. 
It turns out that the set of twisted masses
 in the heterotic model  has $N$ complex mass parameters, rather than $N-1$.
 In the general case, when all mass parameters are nonvanishing, \ntwoo
 supersymmetry is {\em spontaneously} broken already at the classical level.
 If at least one of the above mass parameters vanishes, then \ntwoo is unbroken at the classical level.
 The spontaneous breaking of supersymmetry in this case occurs through nonperturbative effects.

\end{abstract}

\end{titlepage}

\section{Introduction}

The recent discovery of non-Abelian strings 
\cite{HT1,ABEKY,SYmon,HT2}
supported by certain four-dimensional supersymmetric gauge theories 
opened an avenue to understanding of a number of dynamical issues which could not have been addressed
previously. Here we will focus on one particular aspect: models describing low-energy dynamics
on the world sheet of various non-Abelian strings.

The starting point was \cite{ABEKY,SYmon,HT2} \ntwo super-QCD (SQCD) with the gauge group U$(N)$,
$N_f$ massive quark hypermultiplets and the Fayet--Iliopoulos (FI) $\xi$ term for the U(1) factor
(for a review see \cite{SYrev}). If $\xi\gg\Lambda^2$
this bulk theory can be treated quasiclassically. Furthermore, 
for $N_f=N$  critical flux tube solutions exist (BPS-saturated both
at the classical and quantum levels) which, in addition to the conventional
(super)translational moduli are characterized by orientational and superorientational moduli.
Low-energy dynamics of these moduli fields is described by
a two dimensional \ntwot sigma model with the CP$(N-1)$ target space. If the mass terms of 
the quark supermultiplets are different, the \ntwot worldsheet model acquires twisted masses
\cite{twisted}. Still, \ntwot supersymmetry (SUSY) on the world sheet is preserved, which guarantees
complete decoupling of the (super)translational and (super)orientational sectors
of the world-sheet model. The (super)translational sector is presented by a free \ntwot field
theory.

Moving towards \none bulk theories one discovers \cite{Edalati,SYhet,BSYhet}
 a novel class of deformations
of the world-sheet \ntwot supersymmetric CP$(N-1)$ model, currently known as 
{\em heterotic} CP$(N-1)$ model. Assume we deform  basic \ntwo SQCD\,\footnote{
The  gauge group is assumed to be U$(N)$ and we take $N$   quark hypermultiplets
with {\em equal} mass terms.}
 by the superpotential mass term for the adjoint supermultiplet,
\beq
{\mathcal W}_{3+1}=\frac{\mu}{2} \left[{\mathcal A}^2
+  ({\mathcal A}^a)^2\right],
\label{defpo}
\eeq
where $\mu$ is a common mass parameter for the chiral
superfields in \ntwo gauge supermultiplets,
U(1) and SU($N$), respectively. The subscript 3+1 tells us that the deformation 
superpotential (\ref{defpo}) refers to the four-dimensional bulk theory.
Then the \ntwo supersymmetry in the bulk is lost, as it is explicitly broken down to \nonen.
Gone with \ntwo in the bulk is \ntwot SUSY on the world sheet.
Decoupling of the (super)translational and (super)orientational sectors disappears.
Instead, the two fermionic moduli from the supertranslational sector (right-handed spinors)
get connected with the superorientational fermionic moduli.
The corresponding coupling constants are  presented by one complex number \cite{Edalati}
related~\cite{SYhet,BSYhet} to the deformation parameter $\mu$
in Eq.~(\ref{defpo}).
The heterotic coupling of the
supertranslational moduli fields reduces the world-sheet supersymmetry from
\ntwot to \ntwoo$\!\!$.
This is the origin of the alternative name,
 \ntwoo \CPC model. The heterotic CP$(N-1)$ model has rich dynamics (it was solved \cite{SYhetlN}
 at large $N$) leading to the spontaneous breaking of 
 \ntwoo SUSY due to nonperturbative effects.\footnote{The latter statement refers to the particular
 form of the bulk deformation quoted in Eq.~(\ref{defpo}).} 
 
 This heterotic model follows from the bulk theory   described above provided all
 mass terms of the quark hypermultiplets are set equal. If they are unequal,
 it is reasonable to expect that, just as in the \ntwo case,
 the inequality of masses in the bulk will manifest itself  on the world sheet as
 twisted masses. Leaving aside (for future studies)
 derivation of the heterotic CP$(N-1)$ model with twisted masses on the world sheet
 from the \none-deformed microscopic theory in the bulk,
 we will focus in this work on the
  heterotic CP$(N-1)$ model with twisted masses {\em per se}. To the best of our knowledge
  nobody has ever discussed this model. 
  We address two questions, (i) whether or not \ntwoo supersymmetry at the classical level
  survives the introduction of the twisted masses into heterotic CP$(N-1)$,
  and (ii) construction of the corresponding Lagrangian both in the gauged and geometrical formulations.
  The large-$N$ solution of the model will be the next step \cite{BSY5}.
  
  Our findings can be summarized as follows. If the \ntwot model with twisted masses
  contains $N-1$ free mass parameters, its heterotic deformation allows one to introduce two rather than one 
  extra complex parameters. One of them is obvious: 
a complex coupling ($\delta$ or $\tgamma$, see below\footnote{
In this paper we denote the parameter of deformation of the heterotic worldsheet by $\tgamma$ 
which is related to the analogous parameter $\gamma$ originally introduced in \cite{SYhet}, as
$ \tgamma ~=~ \sqrt{2/\beta}\, \gamma $.})
 regulating the strength of the heterotic deformation. The second complex parameter is less obvious,
 being an extra mass parameter. It turns out that the set of twisted masses
 in the heterotic version has $N$ complex mass parameters rather than $N-1$.
 In the general case, when all mass parameters are nonvanishing, \ntwoo
 supersymmetry is {\em spontaneously} broken already at the classical level.
 If at least one of the above mass parameters vanishes the \ntwoo is unbroken at the classical level.
 The spontaneous breaking of SUSY occurs in this case through nonperturbative effects \cite{BSY5}.
 
 The paper is organized as follows. In Section ~\ref{gaugef} we briefly review
the twisted-mass deformed \ntwot CP$(N-1)$ model in the gauge formulation.
 Then we show how one can introduce, additionally, the heterotic deformation of the type discussed above.
 Although \ntwot SUSY is destroyed by the combination of the two deformations,
 \ntwoo is demonstrated to survive.
 We explain why, in addition to the heterotic deformation parameter, an extra mass parameter appears (extra
 compared to the \ntwot model). 
 \newpage
 In Section~\ref{geomf} we follow the same avenue to obtain 
 the Lagrangian of the  twisted-mass deformed heterotic  \ntwot model in the geometric formulation.
 Section \ref{con} summarizes our results and outlines the program of future research in the given direction.

\section{Gauge Formulation}
\label{gaugef}
\setcounter{equation}{0}

In this section we first review two-dimensional \ntwot
CP$(N-1)$ sigma model with twisted masses in the 
gauge formulation and then present its \ntwoo deformation.

\subsection{ \ntwot CP$(N-1)$  model}
\label{ntwotp}

Two-dimensional
supersymmetric \ntwot CP$(N-1)$  model is  known to describe
internal dynamics of  non-Abelian strings in \ntwo
super-QCD with the U$(N)$ gauge group and $N$ flavors
of quarks \cite{HT1,ABEKY,SYmon,HT2}, see
also reviews \cite{Trev,SYrev,Jrev,Trev2}. In the gauge formulation, this model 
with twisted
masses $ m^l $ ($\, l = 1, ...\, N $) is given by the strong coupling limit ($ e^2\to \infty $) of
the following U(1) gauge theory \cite{W79}:
\begin{align}
\label{sigma22}
\notag
 	\mc{L}_\text{(2,2)} & ~~=~~
	\,\frac{1}{4e^2}\,F_{kl}^2  ~+~ \frac{1}{e^2} \left|\p_k \sigma\right|^2 
	~+~ \frac{1}{2e^2}\, D^2
	~+~ \frac{1}{e^2}\, \blar\, i\p_L\, \lar  ~+~  \frac{1}{e^2} \blal\, i\p_R\, \lal
	\\[2mm]
\notag
	&
	\hspace{-0.4cm}
	~+~ 2\beta \biggl \lgroup
	\left| \nabla n \right|^2  ~+~ 2 \Bigl| \sigma - \frac{m^l}
	{\sqrt{2}} \Bigr|^2 \left| n^l \right|^2
	~+~ iD \left( \left|n^l \right|^2 - 1 \right)
	\\
\notag
	&
	\hspace{-0.4cm}\phantom{2\beta\lgroup}
	~+~ \bxir\, i\nabla_L \xir  ~+~ \bxil\, i\nabla_R \xil ~+~
	i\sqrt{2}\, \Bigl( \sigma - \frac{m^l}{\sqrt{2}} \Bigr) \ov{\xi}{}_{Rl} \xi_L^l
	~+~ i\sqrt{2}\, \Bigl( \ov{\sigma} - \frac{\ov{m}{}^l}{\sqrt{2}} \Bigr) \ov{\xi}{}_{Ll} \xi_R^l
	\\[2mm]
	&
	\hspace{-0.4cm}\phantom{2\beta\lgroup}
	~+~ i\sqrt{2}\, \ov{\xi_{[R}\, \lambda}{}_{L]}\, n
	~-~ i\sqrt{2}\, \nbar\,  \lambda_{[R}\, \xi_{L]}
	\biggr\rgroup
	\\[1mm]
\notag
	&
	l~=~1,...\,N,
\end{align}
where
\beq
\nabla_k=\p_k-iA_k, \qquad \nabla_{R,L}=\nabla_0\pm i\nabla_3,
\qquad \lambda_{[R}\, \xi_{L]}= \lambda_R\xi_L-\lambda_L\xi_R\,,
\eeq
while $x_k$ ($k=0,3$) denotes the two coordinates
on the string world sheet. We assume that the string is stretched in the
$x_3$ direction. In the above Lagrangian  $n^l$ are $N$ complex scalar fields of the CP$(N-1)$ model 
and $\xi^l_{R,L}$ are their fermionic superpartners.
All fields of the gauge supermultiplet, namely
the gauge field $A_k$, complex scalar $\sigma$, fermions
$\lambda_{R,L}$ and auxiliary field $D$ are not dynamical
in the  limit $e^2\to\infty$. They can be eliminated
 via algebraic equations of motion. In particular,
integration over $D$ and $\lambda$ give the standard CP$(N-1)$ model
constraints
\beq
|n^l|^2=1, \qquad \bar{n}_l\xi^l_{R,L}=0.
\eeq
Parameters $m^l$ in Eq.~(\ref{sigma22}) are the twisted masses. 

A comment is in order here on our summation conventions for the CP($N-1$) indices $l$, {\it etc.}, since
	they become non-trivial once the twisted masses are introduced.
	The sum in $l$ is always implied if the index is written more than once.
	In the places where this can cause ambiguity we put the summation sign explicitly.
	Furthermore, we specify the range of variation of CP($N-1$) indices
	in the end of equations.
	Finally, we omit the summation sign in those terms where the sum is obvious,
	such as the kinetic terms $ \ov{\xi}\, i\nabla \xi $.

Another comment refers to the twisted masses. From Eq.~(\ref{sigma22})
it is obvious that by virtue of the shift of the $\sigma$ field one can
 always impose an additional condition 
 \begin{equation}
 \sum_{l=1}^N\,\, m^l =0\,.
 \label{adco}
 \end{equation}
 Therefore, in fact, there are $N-1$ independent mass parameters in the \ntwot 
 version of the model.
	
	We will use two normalizations of the physical fields $n$, $\xi$, $\zeta$ in this paper.
	To prove supersymmetry of \eqref{sigma22} it is easier to include the factor of $ 2\beta $ into 
	the big bracket by redefining the corresponding fields,
	so that $ |n|^2 ~=~ 2\beta $.
	However, in order to determine the correspondence between the 
	above model and the geometric formulation of CP$(N-1)$ model with twisted masses it is easier to leave
	it outside, so that $ |n^l|^2 = 1 $.
	
     The model (\ref{sigma22}),
      apart from the two-dimensional Fayet--Iliopoulos (FI) term $ - i D $, is nothing but
        the dimensionally reduced \none four-dimensional SQED.
        From this fact one obtains the following transformation laws of the component fields under 
        \ntwot supersymmetry, with transformation parameters $\epsilon_{R,L}$ and 
      $\bar{\epsilon}_{R,L}$  :
\begin{align}
\notag
  \delta A_{R,L} & ~~=~~ ~~~~~ 2i \lgr  \eerl\, \blarl
                              ~-~ \beerl\, \larl \rgr  , \\
\notag
  \delta\sigma & ~~=~~ -\sqrt{2}
                              \lgr \eer\, \blal ~-~ \beel\,\lar \rgr , 
                              \\
\notag
  \delta\ov{\sigma} & ~~=~~ +\sqrt{2}
                              \lgr \beer\,\lal ~-~ \eel\,\blar \rgr ,
                              \\
\notag
  \delta\lar & ~~=~~ -\,\eer\cdot D ~-~ \frac{1}{2}\,\eer\cdot F_{RL} 
                     ~-~ i\sqrt{2}\, \p_R\sigma \cdot \eel
                     \\
\notag
  \delta\lal & ~~=~~ -\,\eel\cdot D ~+~ \frac{1}{2}\,\eel\cdot F_{RL} 
                     ~-~ i\sqrt{2}\, \p_L\sigma \cdot \eer
                     \\
\notag
  \delta\blar & ~~=~~ -\,\beer\cdot D ~+~ \frac{1}{2}\,\beer\cdot F_{RL} 
                     ~+~ i\sqrt{2}\, \p_R\ov{\sigma} \cdot \beel
                     \\
\notag
  \delta\blal & ~~=~~ -\,\beel\cdot D ~-~ \frac{1}{2}\,\beel\cdot F_{RL}
                     ~+~ i\sqrt{2}\, \p_L\sigma \cdot \beer
                     \\
\notag
  \delta D & ~~=~~ ~~~i\,\eer\, \p_L \blar ~+~ i\,\beer\, \p_L\lar 
                   ~+~ i\,\eel\, \p_R \blal ~+~ i\,\beel\,\p_R\,\lal
                   \\[1mm]
\label{msusy}
  \delta n & ~~=~~ -\,\sqrt{2}\, \epsilon_{[R}\, \xi_{L]} 
                   \\[2mm]
\notag
  \delta\ov{n} & ~~=~~ +\,\sqrt{2}\,\ov{\epsilon_{[R}\, \xi}{}_{L]}
                   \\
\notag
  \delta\xi_R^l & ~~=~~
     -\, i\sqrt{2}\, \beel\, \nabla_R\, n^l ~+~ \sqrt{2}\, \eer\, F^l 
     ~-~ 2i\, \beer\, \Bigl(\sigma - \frac{m^l}{\sqrt{2}}\Bigr)\, n^l
     \\
\notag
  \delta\xi_L^l & ~~=~~
     +\, i\sqrt{2}\, \beer\, \nabla_L\, n^l ~+~ \sqrt{2}\, \eel\, F^l
     ~+~ 2i\, \beel\, \Bigl(\ov{\sigma} - \frac{\ov{m}{}^l}{\sqrt{2}}\Bigr)\, n^l
     \\
\notag
  \delta\ov{\xi}{}_{lR} & ~~=~~
     +\, i\sqrt{2}\, \eel\, \nabla_R\, \ov{n}{}_l 
     ~+~ \sqrt{2}\, \beer\, \ov{F}{}_l 
     ~-~ 2i\, \eer\, \Bigl(\ov{\sigma} - \frac{\ov{m}{}^l}{\sqrt{2}}\Bigr)\, \ov{n}{}_l
     \\
\notag
  \delta\ov{\xi}{}_{lL} & ~~=~~
     -\, i\sqrt{2}\, \eer\, \nabla_L\, \ov{n}{}_l
     ~+~ \sqrt{2}\, \beel\, \ov{F}{}_l
     ~+~ 2i\, \eel\, \Bigl(\sigma - \frac{m^l}{\sqrt{2}}\Bigr)\, \ov{n}{}_l
     \\
\notag
  \delta F^l & ~~=~~
     -\,i\sqrt{2} \lgr \beel\, \nabla_R\, \xi_L^l ~+~ \beer\, \nabla_L \xi_R^l \rgr
     ~-~ 2i\, \ov{\epsilon_{[R}\, \lambda}{}_{L]}\, n^l
     \\
\notag
     & \phantom{~~=~~}
     ~-~ 2i \lgr \beer\, \Bigl(\sigma - \frac{m^l}{\sqrt{2}}\Bigr)\, \xi_L^l 
             ~+~ \beel\, \Bigl(\ov{\sigma} - \frac{\ov{m}{}^l}{\sqrt{2}}\Bigr)\, \xi_R^l \rgr 
     \\
\notag
  \delta \ov{F}{}_l & ~~=~~
     -\,i\sqrt{2} \lgr \eer\, \nabla_L\, \ov{\xi}{}_{lR} ~+~ 
                       \eel\, \nabla_R\, \ov{\xi}{}_{lL} \rgr
     ~+~ 2i\,\ov{n}{}_l\, \epsilon_{[R}\, \lambda_{L]} 
     \\
\notag
     & \phantom{~~=~~}
     ~+~ 2i \lgr \eel \Bigl(\sigma - \frac{m^l}{\sqrt{2}}\Bigr)\, \ov{\xi}{}_{lR} 
             ~+~ \eer \Bigl(\ov{\sigma} - \frac{\ov{m}{}^l}{\sqrt{2}}\Bigr)\, 
                            \ov{\xi}_{lL} \rgr ,
\end{align}
where $F^l$ are $F$ components of the $(n^l,\xi^l)$
supermultiplet, while $F_{RL}=-2i F_{03}$ is a
convenient notation for the gauge field
strength.

	Obviously, with vanishing twisted masses, the theory \eqref{sigma22} is invariant
	under the massless version of the supertransformations \eqref{msusy}.
	The masses themselves can be considered just as constant ``background"
	gauge fields of
	the ``original'' U(1)$^{N-1}$  
	four-dimensional SQED, directed in the $ (x_1,\, x_2) $-plane \cite{HaHo,Dorey}.
	After dimensional reduction they become constant ``$\sigma$''s.
	Hence, supersymmetry should be preserved by twisted mass deformations,\footnote{
The vector supermultiplet $(A_k,\,\sigma,\,\lambda,\,D)$, with only 
          $ \sigma $ nonvanishing and constant, is obviously invariant under \ntwot supersymmetry.}
	and it indeed is \cite{twisted}.

\subsection{ Heterotic \boldmath{\ntwoon} \CPC  model}
\label{hbmnt}

As was mentioned, in   \ntwo supersymmetric bulk theory, the translational
sector of the world-sheet model on the non-Abelian string 
decouples from the orientational one. The translational sector is
associated with the position of the string $x_{0i}$  ($i=1,2$) in the orthogonal plane $(x_1,x_2)$; 
it also includes its fermionic superpartners
$\zeta_L$ and $\zeta_R$. 
The orientational sector is described by the \ntwot CP$(N-1)$ model (\ref{sigma22}). Once \ntwo
breaking deformation (\ref{defpo})
is added in the bulk theory, this decoupling no longer takes place \cite{Edalati}. The translational sector becomes
mixed with the orientational one. In fact, the  fields $x_{0i}$ and 
$\zeta_L$ are still free and do decouple. At the same time,  the 
right-handed translational modulus $\zeta_R$ becomes coupled to
the orientational sector. 

Our next step is to combine the heterotic \ntwoo deformations of massless
\CPC model studied in \cite{Edalati, SYhet,SYhetlN,BSYhet}
with the twisted-mass deformed CP$(N-1)$ model  (\ref{sigma22}).

Let us parenthetically note that, as was shown in \cite{Edalati,SYhet}, 
 the BPS nature of the non-Abelian   string solution is preserved only
if the critical points of the bulk deformation superpotential coincide with the quark masses. 
This is related to the fact that,
if the above condition is fulfilled,  only the  quark scalar fields whose masses are related by \none supersymmetry
to masses of the  gauge bosons are  excited on the string solution.
Other quark fields, with different masses, identically vanish. If the
above condition is not satisfied, other quark fields must be  
non-zero on the solution. This immediately spoils the BPS saturation of the string solution.

Clearly, the only critical point  of the superpotential (\ref{defpo})
is at zero. Therefore, for a generic   choice of nonvanishing
quark mass terms, the above condition is not met. 
Thus, we expect
that the ``BPS-ness" of the non-Abelian string solutions is lost. Below  
  we will see how this four-dimensional
  perspective is translated into the language of the two-dimensional
 world-sheet theory. We will
see that it manifests itself in the spontaneous breaking of 
\ntwoo supersymmetry in the \CPC model on the string world sheet. 
This happens already at the classical level. Note that in the massless case studied in \cite{SYhet,BSYhet} the above 
condition is met, and \ntwoo supersymmetry is preserved
in the world-sheet model at the classical level. Still, it turns
out to be spontaneously broken by quantum (nonperturbative)
effects 
\cite{Tongd,SYhet,SYhetlN}.

In the gauge formulation, the two-dimensional
\ntwoo \CPC  model with twisted
masses  is given by the strong coupling ($ e^2\to\infty $) limit  of
the following U(1) theory
\begin{align}
\label{sigma_full}
\notag
 	\mc{L}_\text{(0,2)} & ~~=~~
	-\,\frac{1}{8e^2}\,F_{RL}^2  ~+~ \frac{1}{e^2} \left|\p_k \sigma\right|^2 
	~+~ \frac{1}{2e^2}\, D^2
	~+~ \frac{1}{e^2}\, \blar\, i\p_L\, \lar  ~+~  \frac{1}{e^2} \blal\, i\p_R\, \lal
	\\[2mm]
\notag
	&
	\hspace{-0.4cm}
	~+~ 2\beta \biggl \lgroup
	\left| \nabla n \right|^2  ~+~ 2 \Bigl| \sigma - 
	\frac{m^l}{\sqrt{2}} \Bigr|^2 \left| n^l \right|^2
	~+~ iD \left( \left|n^l \right|^2 - 1 \right)
	\\
\notag
	&
	\hspace{-0.4cm}\phantom{2\beta\lgroup}
	~+~ \bxir\, i\nabla_L \xir  ~+~ \bxil\, i\nabla_R \xil ~+~
	i\sqrt{2}\, \Bigl( \sigma - \frac{m^l}{\sqrt{2}} \Bigr) \ov{\xi}{}_{Rl} \xi_L^l
	~+~ i\sqrt{2}\, \Bigl( \ov{\sigma} - \frac{\ov{m}{}^l}{\sqrt{2}} \Bigr) \ov{\xi}{}_{Ll} \xi_R^l
	\\[2mm]
	&
	\hspace{-0.4cm}\phantom{2\beta\lgroup}
	~+~ i\sqrt{2}\, \ov{\xi_{[R}\, \lambda}{}_{L]}\, n
	~-~ i\sqrt{2}\, \nbar\,  \lambda_{[R}\, \xi_{L]}
	\\[2mm]
\notag
	&
	\hspace{-0.4cm}\phantom{2\beta\lgroup}
	~+~ \bzr\, i\p_L\, \zr   ~+~   \bff\, \ff
	\\
\notag
	&
	\hspace{-0.4cm}\phantom{2\beta\lgroup}
	~+~ 2i\, \frac{\p^2 \hat{\mc{W}}}{\p \sigma^2}\, \blal\, \zr
	~+~ 2i\, \frac{\p^2 \hat{\ov{\mc{W}}}}{\p \ov{\sigma}{}^2}\, \bzr\, \lal
	~-~ 2i\, \frac{\p \hat{\mc{W}}}{\p \sigma}\, \ff
	~-~ 2i\, \bff\, \frac{\p \hat{\ov{\mc{W}}}}{\p \ov{\sigma}}
	\biggr\rgroup ,
	\\[1mm]
\notag
	&
	l~=~1,...\,N,
\end{align}
	where $ \hat{\mc{W}}(\Sigma) $ is an   \ntwot breaking superpotential 
	function. The hat over ${\mc{W}}$ will remind us that
	this superpotential refers to a two-dimensional theory, rather than to a four-dimensional one.
	In particular, in this paper we consider
$ \hat{\mc{W}}(\Sigma) $ to be   quadratic,
\[
	\hat{\mc{W}}(\Sigma) ~~=~~ \frac{1}{2}\, \delta\, \Sigma^2\,.
\]
The deformation parameter $\delta$ 
was introduced in \cite{SYhet}.\footnote{The relation  $\delta$ of the
world-sheet theory to the  parameter $\mu$
in the four-dimensional  superpotential (\ref{defpo}) was studied in
\cite{SYhet,BSYhet}. To be more exact, this 
relation was derived in \cite{SYhet,BSYhet} for massless
version of the theory. Since 
the mass deformation seems to be  independent of \ntwoo deformation,
 we expect that the same 
relation will hold in the massive theory. The proof of this is
left for  future work. In \cite{SYhet,BSYhet} we obtained that at small $\mu$ the parameter
$
\delta =
 {\rm const}\;\,\frac{g_2^2\mu}{m_W}$,
 while at large $\mu$ we found $\delta =
 {\rm const}\,\frac{\mu}{|\mu|}\;
 \sqrt{\ln{\frac{g^2_2\mu}{m_W}}}$.
 Here $g_2^2$ is the SU($N$) gauge coupling, 
while $m_W$ is the mass of the SU$(N)$ gauge boson of the bulk theory.}

        We prove \ntwoo supersymmetry of the Lagrangian 
        (\ref{sigma_full}) in two steps
	(for now we absorb the factor of $ 2\beta $ into the definition of $ n^l $, $ \xi^l $,
        $ \zeta_R $ and $ \mc{F} $).
        The mass  deformation and the heterotic deformation are independent of each other, as we will shortly prove.
Therefore,   we can consider  first the theory deformed only by the
twisted masses, and then add \ntwoo terms.

As the first step, we discard the superpotential $ \hat{\mc{W}} $.
        Then the theory splits into two decoupled sectors -- orientational and translational. 
        The orientational sector (\ref{sigma22}) was 
        already considered in Section~\ref{ntwotp}.
  The translational sector is free
\[
	\bzr\, i\p_L\, \zr ~~+~~ \bff\,\ff ,
\]
	and invariant under the right-handed supersymmetry \cite{SYhet},
\begin{align}
\notag
        \delta \zr & ~~=~~ ~~~~ \sqrt{2}\, \eer\, \ff\,,
        &
        \delta \bzr & ~~=~~ ~~~~ \sqrt{2}\, \beer\, \bff\,,
        \\
\label{tsusy}
        \delta \ff & ~~=~~ -\, i\sqrt{2}\, \beer\, \p_L \zr\,,
        &
        \delta \bff & ~~=~ -\, i\sqrt{2}\, \eer\, \p_L \bzr
        \,.
\end{align}
	Thus, the direct sum of the two sectors preserves \ntwoo supersymmetry.
	
	The final step is to restore the heterotic deformation $ \hat{\mc{W}} $. 
	The \ntwot fields that mix with the translational sector by means
	of $ \hat{\mc{W}} $ are $ \lambda_L $ and $ \sigma $.
	The supertransformations of the latter do not involve the masses $ m^l $, 
	{\it e.g.} $ \delta\sigma = -\sqrt{2}
                              \left( \eer\, \blal ~-~ \beel\,\lar \right)$, 
                              see Eq.~\eqref{msusy}.
As a result,  the heterotic deformation and the twisted-mass deformation are indeed independent
of each other.

	Finally, now we can assert that the $ \hat{\mc{W}} $ terms are invariant under the overall
	right-handed supersymmetry \eqref{msusy} and \eqref{tsusy}, for  arbitrary 
	superpotential functions $ \hat{\mc{W}}(\sigma) $.
	With the heterotic deformation switched on, $ \hat{\mc{W}}(\sigma) ~\neq~ 0 $,
the shift property is lost in (\ref{sigma_full}):
the shift of $ \sigma $ is no longer a symmetry of the theory. Hence,
there are $ N $ independent twisted mass parameters; physically measurable
quantities depend on
 all of them. 
	In particular, in each Higgs vacuum at weak coupling, $ N - 1 $ parameters define masses of excitations, while one ``extra" parameter determines
	the vacuum energy.

By analogy with massive {\em non}supersymmetric CP$(N-1)$ model
studied in \cite{GSYphtr} in the large-$N$ approximation, we expect
that the model (\ref{sigma_full}) exhibits  two phases, 
separated by a crossover transition,
namely, the week-coupling Higgs phase at large masses and 
the strong-coupling phase at small masses. Detailed study of dynamics
of the heterotic \CPC model (\ref{sigma_full}) is left for future work
\cite{BSY5}. Here we just comment on the week-coupling Higgs phase.

If the twisted masses are large, $|m^l|\gg \Lambda$ (where $\Lambda$ is
the dynamical  scale of the world-sheet theory), the coupling constant
$\beta$ is frozen at the scale of the order of $|m^l|$.
The theory is at weak coupling   and can be studied
perturbatively. We have $N$ vacua. In   each of them,  the vacuum expectation value (VEV) of
$n^l$ is given by
\beq
\langle n^l \rangle =\delta^{ll_0}, \qquad l_0=1,...,N.
\label{higgsn}
\eeq
As follows from (\ref{sigma_full}), in order to find VEV of  the $\sigma$ 
field in the $l_0$-th vacuum, we have to minimize the following potential
\beq
2\left|\sigma -\frac{m^{l_0}}{\sqrt{2}}\right|^2 
+4|\delta|^2\,|\sigma|^2
\label{sigmapot}
\eeq
with respect to $\sigma$.
This minimization gives
\beq
\langle \sigma \rangle = \frac{m^{l_0}}{\sqrt{2}}\,
\frac{1}{1+2|\delta|^2}.
\label{higgssigma}
\eeq
By substituting this back in the potential (\ref{sigma_full})
we get the vacuum energy and masses of all $(N-1)$
elementary fields $n^l$ and $\xi^l$ ($l\neq l_0$). We have,
\begin{align}
\notag
	E_\text{vac}~~\;\, & ~~=~~ |\tgamma|^2\, |m^{l_0}|^2 \,,
	\\[2mm]
\label{hetmass}
	M_\text{ferm}^l\;\,\, & ~~=~~ m^l ~-~ m^{l_0} ~+~ |\tgamma|^2\,m^{l_0} \,,
	\\[2mm]
\notag
	| M_\text{bos}^l |^2 & ~~=~~ | M_\text{ferm}^l |^2 ~-~ |\tgamma|^4\, |m^{l_0}|^2\,,
	\qquad\qquad
	l\neq l_0,
\end{align}
where we introduced a new parameter $\tgamma$ via the relation
\beq
\frac{1}{1+2|\delta|^2}\equiv 1-|\tgamma|^2.
\label{gammadelta}
\eeq

	Although neither the twisted mass deformation, nor the heterotic deformation by themselves break supersymmetry completely, when combined, they lead to the spontaneous \ntwoo supersymmetry breaking
	already at the classical level (unless $m^{l_0}=0$).
 Namely, for non-zero masses 
in each of the Higgs vacua
the vacuum energy does not vanish,  and the boson masses are different from the fermion masses. As was explained
above, this is in accord with our expectations which follow
from the bulk theory picture. In particular, in the special case in which all $N$ masses sit on a circle, 
\[
	m^l ~~=~~ m \cdot e^{2\pi l/N}\,, \qquad\qquad\qquad  l ~=~ 1,...\,, N\,,
\]
	the $ N $ vacua become degenerate.

	Note that supersymmetry restores if one of the masses vanishes.
	The corresponding vacuum becomes supersymmetric
	(at the classical level), as is evident from 
	Eq.~\eqref{hetmass},
\[
	m^{l_0} ~~=~~ 0  \qquad \Rightarrow \qquad E_\text{vac}^{l_0} ~~=~~ 0\,, \qquad\qquad l_0 ~=~ 1,...\,, N\,.
\]
	The theory then becomes a heterotic \CPC model with $ N - 1 $ twisted mass parameters.

	When $ \tgamma = 0 $,   all $ N $ vacua become supersymmetric.
	The heterotic deformation is switched off,
	 and one returns to the twisted-mass deformed \ntwot CP($N-1$) model.
	Although formally there are $ N $ twisted mass parameters, it is well-known that the theory 
	has only $ N - 1 $ physical parameters, more precisely only the mass differences
\[
	M_\text{bos}^i ~~=~~ M_\text{ferm}^i ~~=~~ m^i ~-~ m^{l_0}
\]
enter the spectrum and all other physical
quantities. This can be clearly seen from 
Eq.~(\ref{hetmass}) at $\tgamma =0$. This circumstance is in one-to-one correspondence
with the $\sigma$-shift symmetry of 
the Lagrangian \eqref{sigma22}. Thus, passing
to the heterotic model we acquire not only the heterotic coupling $\delta$ or $\tgamma$,
but, in addition, one ``extra" mass parameter.
	
        To compare the massive heterotic theory in the gauged  formulation to
        the one in the geometric formulation, we will need to eliminate all auxiliary fields from the model. It will be convenient  to have the constraint on $n^i$ in the
        form $ | n^l |^2 = 1 $. To this end we restore the factor 
        $ 2\beta $ in  the model~\eqref{sigma_full}.
        Also we understand that the factor   $ 2\beta $  naturally arises in the 
derivation of  the sigma model from the   string
solution in the microscopic bulk theory.

        We now eliminate the auxiliary fields from \eqref{sigma_full}. 
        As was noted earlier \cite{Edalati}, in the \ntwoo theory the right-handed constraint $ \nbar\,\xir = 0 $ is
changed,
\[
	\nbar\, \xir ~~\propto~~ \ov{\delta}\,.
\]
One can restore the original form of 
the constraint  by performing a shift of the superorientational variable $ \xir $, namely,
\begin{align*}
	& \xi_R' ~~=~~ \xir ~-~ \sqrt{2}\, \ov{\delta}\, n\, \bzr \,, \\[1mm]
	& \ov{\xi}{}_R' ~~=~~ \bxir ~-~ \sqrt{2}\, \delta \nbar\, \zr\,.
\end{align*}
	This obviously changes the normalization of the kinetic term for $ \zr $, which we
	bring back to its canonical form by rescaling $ \zr $,
\[
	\zr ~~\to~~ \frac{1}{1 + 2|\delta|^2}\,\zr ~~\equiv~~ ( 1 - |\tgamma|^2 )\, \zr\,.
\]
As a result of all these transformations,  the following theory emerges\footnote{Note that in 
this paper $ \tgamma ~=~ \sqrt{2/\beta}\,\gamma $, where 
$\gamma$ was introduced in \cite{SYhet}.}:
\begin{align}
\notag
	\frac{\mc{L}}{2\beta} & ~~=~~ \bzr\, i\p_L\, \zr ~+~ 
		| \p n |^2  ~+~ (\nbar\, \p_k n)^2 ~+~ \bxir\, i\p_L\, \xir ~+~ \bxil\, i\p_R\, \xil
	\\[1mm]
\notag
	&~ 
	~-~ (\nbar\, i\p_R n)\, \bxil \xil ~-~  (\nbar\, i\p_L n)\, \bxir \xir
	\\[3mm]
\notag
	&~
	~+~ \tgamma\, i\p_L\nbar \xir\, \zr ~+~ \ov{\tgamma}\, \bxir i\p_L n\, \bzr
	~+~ |\tgamma|^2\, \bxil \xil\, \bzr \zr
	\\[3mm]
\label{sigma_phys}
	&~
	~+~ (1-|\tgamma|^2)\,\bxil\xir\,\bxir\xil ~-~ \bxir\xir\,\bxil\xil
	\\[3mm]
\notag
	&~
	~+~ \sum_l |m^l|^2 \left|n^l \right|^2 
	~-~ i m^l\, \ov{\xi}{}_{Rl} \xi_L^l ~-~ i\ov{m}{}^l\, \ov{\xi}{}_{Ll} \xi_R^l
	\\
\notag
	&~
	~+~ i\tgamma\, m^l\, \ov{n}{}_l \xi_L^l\, \zr 
	~-~ i\ov{\tgamma}\, \ov{m}{}^l\, \ov{\xi}{}_{Ll} n^l\, \bzr
	\\[2mm]
\notag
	&~
	~-~ (1-|\tgamma|^2)
	\lgr \left|\sum m^l |n^l|^2 \right|^2 
		~-~ i m^l\, |n^l|^2 (\bxir\xil) ~-~ i\ov{m}{}^l\, |n^l|^2(\bxil\xir)
	\rgr ,
	\\[2mm]
\notag
	&
	~~~~  l ~=~ 1,...\,N.
\end{align}	
	A few comments are in order concerning Eq.~\eqref{sigma_phys}. 
	Note that in Eq. \eqref{sigma_full} the massive deformation and the heterotic deformation
	were independent from each other, and we used this
	circumstance to prove supersymmetry.
	Now we see that some terms in Eq.~\eqref{sigma_phys}  depend both on $\tgamma$ and
	$m^l$.
	This happens because we have integrated out the auxiliary fields, implying, in turn,  
 that supersymmetry is now realized nonlinearly 
	(see \cite{BSYhet} where supertransformations are written for the
	heterotic CP($N-1$) model).

	With masses set to zero, the model \eqref{sigma_phys} is equivalent to the 
	geometric formulation of the heterotic \ntwoo sigma model \cite{SYhet,BSYhet}.
Our current task is to more closely examine  the mass terms in Eq.~\eqref{sigma_phys}.
	The model still contains redundant fields.
	In particular, there are $N$ bosonic fields $n^l$ and $N$ fermionic $\xi^l$,
	whereas the geometric formulation has $N-1$ corresponding variables, see
	Section~\ref{geomf}.
	We can use the constraints
\[
	\ov{n}{}_l\, n^l ~~=~~ 1\,, \qquad \ov{\xi}{}_l\, n^l ~~=~~ 0
\]
	to get rid of some of them, say $ n^N $ and $ \xi^N $.
	We obtain, for the mass terms,
\begin{align}
\notag
	\frac{\mc{L}}{2\beta} & ~~\supset~~ 
	|\tgamma|^2\, |m^N|^2  
	\\
\notag
	&~
	~+~
	\lgr | m^i - m^N + |\tgamma|^2 m^N |^2 ~-~ |\tgamma|^4\, |m^N|^2 \rgr\, |n^i|^2 
	\\
\notag
	&~
	~-~ (1 - |\tgamma|^2)\, \Bigl|\sum_i\, (m^i - m^N)\,|n^i|^2 \Bigr|^2
	\\
\label{sigma_mass}
	&~
	~-~ i\,(m^i - m^N + |\tgamma|^2 m^N )\,\ov{\xi}{}_{Ri}\,\xi_L^i
	~-~ i\,(\ov{m}{}^i - \ov{m}{}^N + |\tgamma|^2\ov{m}{}^N)\, \ov{\xi}{}_{Li}\, \xi_R^i 
	\\[3mm]
\notag
	&~
	~+~ i\,(1-|\tgamma|^2)\,(m^i - m^N)\,|n^i|^2(\bxir\,\xil) 
	~+~ i\,(1-|\tgamma|^2)\,(\ov{m}{}^i - \ov{m}{}^N)\,|n^i|^2(\bxil\,\xir)
	\\[3mm]
\notag
	&~
	~+~ i \tgamma\, (m^i - m^N)\, \ov{n}{}_i\, \xi_L^i\, \zr
	~-~ i \ov{\tgamma}\, (\ov{m}{}^i - \ov{m}{}^N)\, \ov{\xi}{}_{Li}\, n^i\, \bzr
	\\[3mm]
\notag
	&~
	~-~ i |\tgamma|^2\, m^N\, \ov{\xi}{}_{RN}\, \xi_L^N
	~-~ i |\tgamma|^2\, \ov{m}^N\, \ov{\xi}{}_{LN}\, \xi_R^N\,,
	\qquad\qquad
	i ~=~ 1,...\,N-1\,,
\end{align}
	where we denote
\[
	(\ov{\xi}\, \xi) ~~=~~ \ov{\xi}{}_i\, \xi^i  ~+~  \ov{\xi}{}_N\, \xi^N\,,
	\qquad\qquad\qquad
	i ~=~ 1,...\,N-1\,.
\]

\vspace{2mm}

Note, that at large values of $m^l$ all $N$ Higgs vacua (\ref{higgsn}) of the theory  are
still present in the potential (\ref{sigma_mass}).
One of them, with $l_0=N$, in which  $n^i=0$ for {$i=1,...(N-1)$},
is easily seen from (\ref{sigma_mass}). The vacuum energy and the
masses of the fermion and boson elementary excitations match
expression (\ref{hetmass}) for $l_0=N$. Other $(N-1)$
vacua are still present, but not-so-easy to see from (\ref{sigma_mass}).
They are located at $n^{l_0}=1$ where $l_0=1,...(N-1)$ and  all $  n^i$ with $i\neq l_0$
vanish.

One can easily see these vacua from different equivalent formulations of the theory which emerge if  we choose to eliminate the
field $n^{l_0}$ rather then $n^N$. We stress, however, that
all $N$ vacua are, in principle, seen from any of these
equivalent formulations, and supersymmetry breaking is spontaneous
rather than explicit.

\section{Geometric Formulation}
\label{geomf}	
\setcounter{equation}{0}
	
	The \ntwoo supersymmetric CP($N-1$) model   with heterotic and twisted-mass deformations has  a geometric description which combines elements of
	the  description of the twisted-mass deformed
	\ntwot model (see e.g. \cite{SVZw}) and that of the \ntwoo massless model
	\cite{SYhet}.
	
To begin with let us consider the \ntwot CP($N-1$) model.
In the gauge formulation of this model one has two sets of $ N - 1 $ (anti)chiral 
	superfields $ \Phi^i $ and $ \ov{\Phi}{}^\bj $ ($ i, \bj = 1,..., N-1 $), 
	the lowest components $ \phi^i $, $ \ov{\phi}^\bj $ of which parametrize the target K\"{a}hler
	manifold.
The Lagrangian of the CP($N-1$) model is given by the following sigma model
\[
	\mc{L}
	  ~~=~~ \int\, d^4\theta\, K(\Phi,\ov{\Phi}) ~~=~~ g_{i\bj}\,\p_\mu \phi^i \p_\mu\ov{\phi}{}^\bj
		~+~ \frac{1}{2}\, g_{i\bj}\, \psi^i\, i\overleftrightarrow{\slashed{\nabla}} \ov{\psi}{}^\bj 
		~+~ \frac{1}{4}\, R_{ij\bk\bl}\, \psi^i\psi^j \ov{\psi}{}^\bk \ov{\psi}{}^\bl~,
\]
	where $ K(\phi,\ov{\phi}) $ is the K\"ahler potential, 
	$ g_{i\bj} $ is its K\"ahler metric
\[
	g_{i\bj} ~~=~~ \frac{\p^2 K}{\p\phi^i\,{\p\ov{\phi}{}^\bj}}~,
	\qquad\qquad
	g^{i\bk} ~~=~~ \left(g^{-1}\right)^{\bk i}~,
\]
	$ \nabla_\mu $ is  the (target space) covariant derivative,
\begin{align}
\notag
	(\nabla_\mu \ov{\psi})^\bj & ~~=~~ \left\{ \p_\mu \delta^\bj_{\ \bm} ~+~
						\Gamma_{\bm\bk}^\bj\, \p_\mu(\ov{\phi}{}^\bk) \right\} \ov{\psi}{}^\bm~,
	& \Gamma^\bi_{\bk\bl} & ~~=~~ g^{m\bi}\,\p_\bl\, g_{m\bk}~,
	\\[3mm]
\label{covd}
	(\psi \overleftarrow{\nabla}{}_\mu)^i & ~~=~~ 
			\psi^m \left\{ \overleftarrow{\p}{}_\mu\delta^i_{\ m} ~+~
						\Gamma^i_{mk}\, \p_\mu(\phi^k) \right\}~,
	& \Gamma^i_{kl} & ~~=~~ g^{i\bm}\, \p_l\, g_{k\bm}~,
\end{align}
	and $ R_{ij\bk\bl} $ is the Riemann tensor 
\[
	R_{ij\bk\bl} ~~=~~ \p_i\,\p_\bk\, g_{j\bl} ~-~ g^{m\bm}\; \p_i\, g_{j\bm}\, \p_\bk g_{m\bl}~.
\]
	For the CP($N-1$) model one chooses the K\"ahler potential in the following way:
\[
	K(\Phi, \ov{\Phi}) ~~=~~ \ln \lgr 1 ~+~ \ov{\Phi}{}^\bj \delta_{\bj i} \Phi^i \rgr \,,
\]
	which corresponds to the Fubini--Study metric,
\[
	g_{i\bj} ~~=~~ \frac{1}{\chi}\lgr  \delta_{i\bj} ~-~ \frac{1}{\chi}
				  \delta_{i\bi}\,\ov{\phi}^\bi\, \delta_{j\bj}\,\phi^j \rgr,
	\qquad\qquad \text{where~~}
	\chi ~~=~~ 1 ~+~ \ov{\phi}{}^\bj \delta_{\bj i} \phi^i~.
\]
	In this case,
\[
	\Gamma^\bi_{\bk\bl} ~~=~~ -\, \frac{\delta^\bi_{\ (\bk} \delta_{\bl) i}\, \phi^i}{\chi}\,,  
	\qquad\qquad 
	\Gamma^i_{kl} ~~=~~ -\, \frac{\delta^i_{\ (k} \delta_{l)\bi}\,\ov{\phi}{}^\bi}{\chi}\,,
\]
	and the Riemann tensor takes the form
\[
	R_{ij\bk\bl} ~~=~~ -\,g_{i(\bk}\,g_{\bl)j}~.
\]
	
	As was shown in \cite{SYhet},  the \ntwoo deformation of the CP($N-1$) model can be obtained by
	introduction of the right-handed supertranslational modulus $ \zeta_R $ via a ``right-handed'' 
	supermultiplet $ \mc{B} $,
\begin{align*}
	\mc{B} & ~~=~~ \lgr \zr ~+~ \sqrt{2}\,\theta_R\mc{F} \rgr \ov{\theta}{}_L~, \\[2mm]
	\ov{\mc{B}} & ~~=~~ \theta_L \lgr \bzr ~+~ \sqrt{2}\, \ov{\theta}{}_R \ov{\mc{F}} \rgr.
\end{align*}
	The latter expressions describe superfields covariant only under the right-handed supersymmetry, 
	while explicitly breaking the left-handed one.
	In a sense, $ \mc{B} $ is an analog of the \ntwoo supermultiplet $\Xi$ in the two-dimensional 
	superfield formalism \cite{Edalati} --
	the above supermultiplet containing only one physical field, which is the supertranslational
	fermionic variable. ${\mathcal F}$ is the auxiliary component of the $\mc{B}$ superfield.
	The distinction is that $ \mc{B} $ happens to be a twisted superfield,
\[
	D_R \mc{B} ~~=~~ \ov{D}_L\mc{B} ~~=~~ 0\,.
\]
One then constructs the heterotic Lagrangian
\beq
\label{exte}
	\mc{L}_{(0,2)} ~~=~~ \frac{2}{g_0^2} \int\, d^4\theta\, \lgr K(\Phi,\ov{\Phi}) 
		~-~ 2\, \ov{\mc{B}}\,\mc{B}  
		~+~  \sqrt{2}\, \tgamma\,\mc{B}\,K  ~+~ \sqrt{2}\, \ov{\tgamma}\,\ov{\mc{B}}\,\ov{K} \rgr,
\eeq
	which obviously respects the invariance on the target space CP($N-1$).
	Here
\[	
	\frac{2}{g_0^2} ~~=~~ 2 \beta
\]
	is the coupling constant of the sigma model.
	The second term in Eq.~\eqref{exte} generates the kinetic term for $ \zr $, while the last two terms 
	are responsible for the mixing between $ \zr $ and $ \xi_{R,L} $.
	Explicitly, in components, one has,
\begin{align}
\notag
	\frac{\mc{L}_{(0,2)}}{2\beta} & ~~=~~  \bzr\, i\p_L\, \zr 
			~+~ g_{i\bj}\,\p_\mu \phi^i \p_\mu\ov{\phi}{}^\bj
			~+~ \frac{1}{2}\, g_{i\bj}\, \psi^i_R\, i\overleftrightarrow{\nabla}{}_{\!\!L} \ov{\psi}{}^\bj_R 
			~+~ \frac{1}{2}\, g_{i\bj}\, \psi^i_L\, i\overleftrightarrow{\nabla}{}_{\!\!R} \ov{\psi}{}^\bj_L 
	\\[3mm]
\label{cpn-1g}
			& 
			~~+~~ \tgamma\, g_{i\bj}\, (i \p_L \ov{\phi}{}^\bj)\, \psi_R^i\, \zr
			~+~ \ov{\tgamma}\, g_{i\bj}\, \ov{\psi}{}_R^\bj (i \p_L \phi^i)\, \bzr
			~+~ |\tgamma|^2\, \bzr\,\zr \cdot ( g_{i\bj}\, \ov{\psi}{}_L^\bj\, \psi_L^i )
	\\[3mm]
\notag
			& 
			~~+~~ (1 \!-\! |\tgamma|^2)\, (g_{i\bk}\, \ov{\psi}{}_R^\bk\, \psi_L^i)\,
						     (g_{j\bl}\, \ov{\psi}{}_L^\bl\, \psi_R^j)
			~-~ (g_{i\bk}\, \ov{\psi}{}_R^\bk\, \psi_R^i)\, (g_{j\bl}\, \ov{\psi}{}_L^\bl\, \psi_L^j)~.
\end{align}
	This action was originally introduced in \cite{SYhet}, although in a slightly different
	normalization.
	The two actions match if one normalizes the fermions $ \zr $ canonically, and
	takes into account that our deformation parameter $ \tgamma $ is related to $ \gamma $ 
	of Ref.~\cite{SYhet} as
\[
	\tgamma ~~=~~ \sqrt{2} g_0\cdot \gamma\,.
\]
	
	The geometric form \eqref{cpn-1g} can be related to the gauge formulation (Eq.~\eqref{sigma_phys} with
	vanishing twisted masses) via the following stereographic projection
\begin{align}
\notag
	n^i & ~~=~~ \frac{\phi^i}{\sqrt{\chi}}~,
	& 
	\ov{n}_\bi & ~~=~~ \frac{\ov{\phi}{}^\bi}{\sqrt{\chi}}~,
\\[2mm]
\label{stereo}
	n^N & ~~=~~ \frac{1}{\sqrt{\chi}}~,
	& n^N & ~~\in~~ \mc{R}~,
\\[2mm]
\notag
	\xi^i & ~~=~~ \frac{1}{\sqrt{\chi}} \lgr \psi^i ~-~ \frac{(\ov{\phi}\psi)}{\chi}\,\phi^i \rgr,
	& 
	\ov{\xi}{}_\bi & ~~=~~ \frac{1}{\sqrt{\chi}} 
					\lgr \ov{\psi}{}^\bi ~-~ \frac{(\ov{\psi}\phi)}{\chi}\, \ov{\phi}{}^\bi \rgr,
\\[2mm]
\notag
	\xi^N & ~~=~~ -\, \frac{(\ov{\phi} \psi)}{\chi^{3/2}}~,
	&
	\ov{\xi}{}_N & ~~=~~ -\, \frac{(\ov{\psi} \phi)}{\chi^{3/2}}~,
\end{align}
	where $	i,\, \bi ~=~ 1, ..., N-1 $ and we shortcut the contractions 
	$ (\ov{\psi} \phi) ~=~ \delta_{i\bj}\, \ov{\psi}{}^\bj \phi^i $.
	Here we chose $ n^N $ to be real given an overall phase freedom of the CP($N-1$) variables $ n^l $.
	Also we singled out $ n^N $ to be special and equal to $ 1/\sqrt{\chi} $, 
	which corresponds to picking out one of the $ N $ vacua. Indeed,
	the representation (\ref{stereo}) is very convenient for
	analyzing the vacuum lying at $\phi^i=0$ $(i=1,2, ..., N-1)$. In this representation all other $N-1$ vacua do not disappear, 
	but they lie at infinity in the $\phi^1, ... \phi^{N-1}$ parametrization of the target space.
Needless to say, the role of $n^N$ in (\ref{stereo}) can be assumed by any other
$n^{l_0}$  $(l_0 = 1,2, ..., N-1)$.	Then the $l_0$-th vacuum will be easily accessible, while
the $N$-th one will move to infinity.

\subsection{Twisted Masses}
\label{twim}

	The twisted-mass deformation is carried out by formally lifting the theory to  four-dimensional space and introducing a set of
four-dimensional vector superfields $ V^i $,
\begin{equation}
\label{Vi}
	V^i ~~=~~ A_1^i\, \theta\sigma_1\ov{\theta} ~+~ A_2^i\, \theta\sigma_2\ov{\theta}\,,
\end{equation}
	with only the ``transverse'' components of the gauge field nonvanishing 
	(remember, we assume that the string is ``oriented'' in the ($x_0$, $x_3$)-plane), and with $ \lambda $ and $ D $ equal to zero ($ \lambda $ and $ D $ are other components of the vector superfield).
	The  components $A_1^i$ and $A_2^i$ are constant and define the twisted masses\footnote{We note that relations \eqref{mGi} assume 
	the $ \ov{\Phi}\, e^{\displaystyle V}\, \Phi $ normalization of the gauge field $ V $ inside the kinetic term. 
	In the case when the normalization $ \ov{\Phi}\, e^{\displaystyle - 2 V} \Phi $ is used, 
	as we do in the dimensional reduction of SQED in Section~\ref{gaugef},
	the relations instead become $ m^k ~=~ A_1^k ~+~ i\, A_2^k $\,.}
	through the following relations:
\begin{equation}
\label{mGi}
	m_G^k ~~=~~ -\, \frac{A_1^k ~+~ i\, A_2^k}{2}\,,
	\qquad\qquad k ~=~ 1, ...\, N-1\,.
\end{equation}
	These vector superfields are precisely the same kind of superfields that could give the twisted masses $m^l$
in the model \eqref{sigma_full},
	see discussion after Eq.~\eqref{msusy}. However,  now their number   is $ N - 1 $ instead of $ N $.
	In particular, as was mentioned, these superfields preserve \ntwot supersymmetry
	after dimensional reduction to two dimensions.
	Until Section~\ref{sbreaking} we will not  dwell on the obvious fact that the number of the
	mass parameters in the geometric formulation so far is less (by one) than that of the gauge formulation.
	
	The theory is then gauged with the above vector fields,
\[
	K(\Phi^i, \ov{\Phi}{}^\bj\, V^i) ~~=~~
		\ln \lgr 1 ~+~ \ov{\Phi}{}^\bj\, \delta_{i\bj}\, e^{V^i} \Phi^i \rgr ,
\]	
	with the same action as in Eq.~\eqref{exte},
\beq
\label{Ltw}
	\mc{L}_{(0,2)}^\text{tw.m.} ~~=~~ \frac{2}{g_0^2} \int\, d^4\theta\, \lgr K(\Phi,\ov{\Phi}, V^i) 
		~-~ 2\, \ov{\mc{B}}\,\mc{B}  
		~+~  \sqrt{2}\, \tgamma\,\mc{B}\,K  ~+~ \sqrt{2}\, \ov{\tgamma}\,\ov{\mc{B}}\,\ov{K} \rgr.
\eeq
	The action of the theory \eqref{Ltw} can be calculated by introducing covariantly-chiral superfields
\begin{align*}
	X^i & ~~=~~ \Phi^i\,, \\
	\ov{X}{}^\bj & ~~=~~ e^{V^\bj} \ov{\Phi}{}^\bj\,,
\end{align*}
	in terms of which the K\"{a}hler potential takes the original form
\[
	K ~~=~~ \ln \lgr 1 ~+~ \ov{X}{}^i X^i \rgr\,.
\]
	It turns out that in the above integral one can freely replace $ D_\alpha $ and $ \ov{D}{}_\alpha $ with
	covariant $ \nabla^{(\bj)}_\alpha $ and $ \ov{\nabla}{}^{(i)}_\alpha $ at any convenient occurrence.
	This makes calculation of \eqref{Ltw} straightforward, and the only obvious difference with the massless
	case comes from the algebra of the covariant derivatives $ \nabla^{(\bj)}_\alpha $ and 
	$ \ov{\nabla}{}^{(i)}_\alpha $, {\it i.e.} from the presence of the constant gauge fields.
	For calculation of the conjugate term $ \sqrt{2}\, \ov{\tgamma}\,\ov{\mc{B}}\,\ov{K} $ one can find convenient to 
	use covariantly-antichiral variables
\begin{align*}
	Y^i & ~~=~~ e^{V^i} \Phi^i\,, \\
	Y^\bj & ~~=~~ \ov{\Phi}{}^\bj\,.
\end{align*}
	Needless to say that the result is obtained from the massless theory by the ``elongation" 
	of the space-time derivatives (however, formally, in four dimensions). Namely, in this way we arrive at
\begin{align}
\notag
	\frac{\mc{L}_{(0,2)}^\text{tw.m.}}{2\beta} & ~~=~~ 
	g_{i\bj}\,(\nabla_\mu \phi^i)\, (\nabla_\mu\ov{\phi}{}^\bj)
	~+~ \frac{1}{2}\, g_{i\bj}\, \psi^i\, i\overleftrightarrow{\slashed{\nabla}} \ov{\psi}{}^\bj
	\\
\notag
	&~
	~-~ \p_i\p_\bk g_{l\bj}\, \psi^i_{R}\, \psi^l_{L}\, \ov{\psi^\bk_{R}\, \psi}{}^\bj_{L}
	~+~ g_{i\bj}\, F^i\,\ov{F}{}^\bj
	~-~ \p_\bk g_{i\bj}\, F^i\, \ov{\psi_R^\bk\, \psi}{}_L^\bj
	~+~ \p_i g_{l\bj}\, \psi_R^i\, \psi_L^l\, \ov{F}{}^\bj
	\\[3mm]
\notag
	&~
	~+~ \bzr\, i\p_L\, \zr ~+~ \bff \ff 
	~+~ \tgamma\, \ff\, g_{i\bj}\, \psi_R^i\, \bpsi_L^\bj 
	~+~ \btgamma\, \bff\, g_{i\bj}\, \psi_L^i\, \bpsi_R^\bj
	\\[1mm]
\label{geomlin}
	&~
	~-~ \tgamma\, \ff\, \frac{1}{\chi}\,\phi^\bj\, (\nabla_1^\bj + i \nabla_2^\bj) \bphi^\bj
	~+~ \btgamma\, \bff\, \frac{1}{\chi}\,\bphi^\bi\, (\nabla_1^i - i \nabla_2^i) \phi^i
	\\[1mm]
\notag
	&~
	~-~ \tgamma\,\zr\cdot g_{i\bj}\, F^i\, \bpsi_L^\bj
	~+~ \btgamma\, \bzr\cdot g_{i\bj}\, \ov{F}{}^\bj\, \psi_L^i
	\\[3mm]
\notag
	&~
	~-~ \tgamma\, \zr\, \p_i g_{k\bj}\, \psi_R^i\, \psi_L^k \cdot \bpsi_L^\bj
	~-~ \btgamma\, \bzr\, \p_\bk g_{i\bj}\, \ov{\psi_R^\bj\, \psi}{}_L^\bk \cdot \psi_L^i
	\\[3mm]
\notag
	&~
	~-~ \tgamma\, \zr\, g_{i\bj}\, \psi_R^i\, i\p_L \bphi^\bj
	~-~ \btgamma\, \bzr\, g_{i\bj}\, \bpsi_R^\bj\, i\p_L \phi^i
	\\[3mm]
\notag
	&~
	~-~ \tgamma\, \zr\, g_{i\bj}\, \psi_L^i\, (\nabla_1^\bj + i\nabla_2^\bj) \bphi^\bj
	~-~ \btgamma\, \bzr\, g_{i\bj}\, \bpsi_L^\bj\, (\nabla_1^i - i\nabla_2^i) \phi^i
	\,.
\end{align}
	The mass terms here are hidden in the covariant derivatives
\begin{align*}
	\nabla_\mu^i & ~~=~~ \p_\mu ~+~ \frac{i}{2}\,A_\mu^i\,,  \\[3mm]
	\nabla_\mu^\bj & ~~=~~ \p_\mu ~-~ \frac{i}{2}\,A_\mu^\bj \,,
	&& \!\!\!\!\!\!\! \mu ~=~ 0,...\,3\,,
	\\[3mm]
	(\nabla_\mu \ov{\psi})^\bj & ~~=~~ \left\{ \nabla_\mu^\bm \delta^\bj_{\ \bm} ~+~
						\Gamma_{\bm\bk}^\bj\, \nabla_\mu^\bk (\ov{\phi}{}^\bk) \right\} \ov{\psi}{}^\bm~,
	& \Gamma^\bi_{\bk\bl} & ~~=~~ g^{m\bi}\,\p_\bl\, g_{m\bk}~,
	\\[3mm]
	(\psi \overleftarrow{\nabla}{}_\mu)^i & ~~=~~ 
			\psi^m \left\{ \overleftarrow{\nabla}{}_\mu^m \delta^i_{\ m} ~+~
						\Gamma^i_{mk}\, \nabla_\mu^k (\phi^k) \right\}~,
	& \Gamma^i_{kl} & ~~=~~ g^{i\bm}\, \p_l\, g_{k\bm}\,.
\end{align*}
	Although the space-time index $\mu$ formally runs over all four values, we understand
	that the derivatives $ \p_\mu $ with respect to the transverse coordinates 
	$(\mu=1,2$) should be ignored in our two-dimensional
	theory.

Next, we need to exclude the auxiliary fields $ F^i $ and $ \ff $, since they have no analogs
	in the gauge formulation of the theory 
	(more precisely, although we did introduce  exactly the same $\mc{B}$ superfield into the gauge formulation, 
	the highest component $\ff$ of that multiplet played a different role in Eq.~\eqref{sigma_full} than in Eq.~\eqref{geomlin}).
	Also, we substitute the masses, noting that the covariant derivatives in Eq.~\eqref{geomlin} 
	enter in convenient combinations, 
\begin{align*}
	\nabla_1^\bj ~+~ i\,\nabla_2^\bj & ~~=~~ \phantom{-} i\,m_G^\bj\,,    \\[2mm]
	\nabla_1^i ~-~ i\,\nabla_2^i     & ~~=~~ -i\,\ov{m}{}_G^i\,,
	\qquad\qquad\qquad i,\,\bj ~=~ 1,...\,N-1\,.
\end{align*}
The $ F^i $-term conditions are the same as in the massless heterotic theory, while the $ \ff $-term condition
	gets modified by the masses,
\beq
\label{ffterm}
	\ff ~~=~~ -\, \btgamma\, g_{i\bj}\, \psi_L^i\, \bpsi_R^\bj
	      ~+~ i\, \btgamma\, \ov{m}_G^i \frac{\bphi^i\, \phi^i}{\chi}\,.
\eeq
	As a result, we obtain, 
\begin{align}
\notag
	\frac{\mc{L}_{(0,2)}^\text{tw.m.}}{2\beta} & ~~=~~ 
	\bzr\, i\p_L\, \zr \\[1mm]
\notag
	&\!\!\!\!\!\!\!\!\!\!\!\!
	~+~ g_{i\bj}\, \p_\mu\phi^i\, \p_\mu\bphi^\bj ~+~ g_{i\bj}\, m_{G\mu}^i\,m_{G\mu}^\bj\, \phi^i\, \bphi^\bj
	\\[2mm]
\notag
	&\!\!\!\!\!\!\!\!\!\!\!\!
	~+~ \frac{1}{2}\,g_{i\bj}\,\psi_R^i\,i\overleftrightarrow{\nabla}{}_{\!\!L}^{(0)} \bpsi_R^\bj 
	~+~ \frac{1}{2}\,g_{i\bj}\,\psi_L^i\,i\overleftrightarrow{\nabla}{}_{\!\!R}^{(0)} \bpsi_L^\bj 
	~+~ \frac{i}{2}\,g_{i\bj}\,\psi_L^i\, \overleftrightarrow{m}{}_{\!G} \bpsi_R^\bj
	~+~ \frac{i}{2}\,g_{i\bj}\,\psi_R^i\, \overleftrightarrow{\ov{m}}{}_{\!G} \bpsi_L^\bj
	\\[2mm]
\label{sgeom_phys}
	&\!\!\!\!\!\!\!\!\!\!\!\!
	~-~ (g_{i\bj}\, \psi_R^i\, \bpsi_R^\bj)\,(g_{k\bl}\, \psi_L^k\, \bpsi_L^\bl) 
	~+~ (1 \!-\!|\tgamma|^2)(g_{i\bj}\, \psi_R^i\, \bpsi_L^\bj)\,(g_{k\bl}\,\psi_L^k\,\bpsi_R^\bl)
	\\[3mm]
\notag
	&\!\!\!\!\!\!\!\!\!\!\!\!
	~+~ \tgamma\, g_{i\bj}\, i\p_L\bphi^\bj\, \psi_R^i\,\zr 
	~+~ \btgamma\, g_{i\bj}\, \bpsi_R^\bj\, i\p_L\phi^i\, \bzr
	~+~ i\tgamma\, g_{i\bj}\, m_G^\bj\,\bphi^\bj\, \psi_L^i\,\zr
	~-~ i\btgamma\, g_{i\bj}\,\ov{m}{}_G^i\, \bpsi_L^\bj\, \phi^i\,\bzr
	\\[4mm]
\notag
	&\!\!\!\!\!\!\!\!\!\!\!\!
	~+~ |\tgamma|^2\, g_{i\bj}\, \bpsi_L^\bj\, \psi_L^i\, \bzr\, \zr
	\\[2mm]
\notag
	&\!\!\!\!\!\!\!\!\!\!\!\!
	~+~ i\, |\tgamma|^2\, (g_{i\bj}\,\psi_L^i\,\bpsi_R^\bj)\, \cdot m_G^k \frac{\bphi^k\, \phi^k}{\chi}
	~+~ i\, |\tgamma|^2\, (g_{i\bj}\,\psi_R^i\,\bpsi_L^\bj)\, \cdot \ov{m}{}_G^\bk \frac{\bphi^\bk\,\phi^\bk}{\chi}
	\\[2mm]
\notag
	&\!\!\!\!\!\!\!\!\!\!\!\!
	~+~ |\tgamma|^2 \cdot \ov{m}{}_G^\bj\,\frac{\bphi^\bj\,\phi^\bj}{\chi}
			\cdot m_G^i\, \frac{\bphi^i\, \phi^i}{\chi}\,.
\end{align}
	Some comments are in order here on the notation used in this expression. First,
	  $ \nabla^{(0)} $ denotes the nongauge (but still covariant) derivative \eqref{covd}
	of the massless theory.
	The index $ \mu ~=~ 1,\, 2$ of the masses $ m_G^i $ denotes their real and imaginary parts, respectively,
	which is consistent with Eq.~\eqref{mGi}.
	Finally, the matrix $ \overleftrightarrow{m}{}_{\!G} $ acts on   spinors in accordance with
	the following formula:
\begin{align*}
	(m_G\, \bpsi_{R,L})^\bj & ~~=~~ 
		\bigl( m_G^\bm\, \delta_{\bm}^{~\bj} ~+~ m_G^\bk\, \Gamma^\bj_{\bm\bk}\, \bphi^\bk \bigr)\, \bpsi_{R,L}^\bm\,,
	\\[2mm]
	(\psi_{R,L}\,\overleftarrow{m}{}_{\!G})^i & ~~=~~
		-\, \psi_{R,L}^m\, \bigl( m_G^m\, \delta_m^{~i} ~+~ m_G^k\, \Gamma^i_{mk}\, \phi^k \bigr)\,,
	\\[2mm]
	\overleftrightarrow{m}{}_{\!G} & ~~=~~ \overrightarrow{m}{}_{\!G} ~-~ \overleftarrow{m}{}_{\!G}\,,
\end{align*}
	with the identical prescription for the conjugate mass matrix $ \overleftrightarrow{\ov{m}}{}_{\!G} $.

	It is instructive to compare the theory \eqref{sgeom_phys} with the gauge formulation of  the
	twisted-mass deformed heterotic sigma
	model  \eqref{sigma_phys}. To this end one
can use the (inverted) correspondence rules \eqref{stereo}. 
	For the massless theory, this job was done in \cite{BSYhet}, where it was shown that one formulation exactly
	matches onto the other.
	Therefore, we need to prove the correspondence only for the mass  terms.
	We have
\begin{align}
\notag
	\frac{\mc{L}_{(0,2)}^\text{tw.m.}}{2\beta} & ~~\supset~~ 
	g_{i\bj}\, m_{G\mu}^i\,m_{G\mu}^\bj\, \phi^i\, \bphi^\bj
	~+~ i\,\frac{1}{2}\,g_{i\bj}\,\psi_L^i\, \overleftrightarrow{m}{}_{\!G} \bpsi_R^\bj
	~+~ i\,\frac{1}{2}\,g_{i\bj}\,\psi_R^i\, \overleftrightarrow{\ov{m}}{}_{\!G} \bpsi_L^\bj
	\\[2mm]
\notag
	&~
	~+~ i\tgamma\, g_{i\bj}\, m_G^\bj\,\bphi^\bj\, \psi_L^i\,\zr
	~-~ i\btgamma\, g_{i\bj}\,\ov{m}{}_G^i\, \bpsi_L^\bj\, \phi^i\,\bzr
	\\[2mm]
\notag
	&~
	~+~ i\, |\tgamma|^2\, (g_{i\bj}\,\psi_L^i\,\bpsi_R^\bj)\, \cdot m_G^k \frac{\bphi^k\, \phi^k}{\chi}
	~+~ i\, |\tgamma|^2\, (g_{i\bj}\,\psi_R^i\,\bpsi_L^\bj)\, \cdot \ov{m}{}_G^\bk \frac{\bphi^\bk\,\phi^\bk}{\chi}
	\\[1mm]
\notag
	&~
	~+~ |\tgamma|^2 \cdot \ov{m}{}_G^\bj\,\frac{\bphi^\bj\,\phi^\bj}{\chi}
			\cdot m_G^i\, \frac{\bphi^i\, \phi^i}{\chi}
	~~=~~
	\\[1mm]
\label{sgeom_mass}
	&\!\!\!\!\!\!\!\!
	~=~
	\sum_i |m_G^i|^2 \left|n^i \right|^2 ~-~ (1 \!-\! |\tgamma|^2)\, \Bigl| \sum_i m_G^i\, |n^i|^2 \Bigr|^2 
	\\
\notag
	&~
	~-~ i\,m_G^i\,\bxi_{Ri}\,\xi_L^i 
	~-~ i\,\ov{m}{}_G^i\, \bxi_{Li}\, \xi_R^i ~+~
	\\[4mm]
\notag
	&~
	~+~ i\,(1\!-\!|\tgamma|^2)\, m_G^i\,|n^i|^2\, (\bxir\,\xil) 
	~+~ i\,(1\!-\!|\tgamma|^2)\, \ov{m}{}_G^i\,|n^i|^2\, (\bxil\,\xir)
	\\[3mm]
\notag
	&~
	~+~ i\,\tgamma\, m_G^i\,\ov{n}{}_i\, \xi_L^i\, \zr
	~-~ i\,\btgamma\, \ov{m}{}_G^i\, \bxi_{Li}\, n^i\, \bzr\,,
	\\[3mm]
\notag
	&~~~
	\text{$i$, $\bj$, $k$, $\bk$} ~=~ 1,...\,, N-1\,.
\end{align}
	Here 
$
	(\ov{\xi}\, \xi) ~~=~~ \ov{\xi}{}_i\, \xi^i  ~+~  \ov{\xi}{}_N\, \xi^N
$,
	similarly to Eq.~\eqref{sigma_mass}.
	Comparing Eq.~\eqref{sgeom_mass} with Eq.~\eqref{sigma_mass}, we   see that the former
	does not   exactly match   the latter.
	It would match if we set $ m^N = 0 $ in the gauge formulation.
	As was discussed in Section~\ref{gaugef}, this would be a heterotic CP($N-1$) sigma model
	with $ N-1 $ twisted mass parameters, which has a supersymmetric vacuum.
	Eq.~\eqref{sgeom_phys} describes excitations around this vacuum.

	That there was a problem with the number of twisted mass parameters in the geometric formulation 
	was obvious from the beginning, see Eq.~\eqref{mGi}.
	The number of physical fields is the same in both formulations, but the number of masses is not.
	Not only that, the theory \eqref{sgeom_phys} will always be (classically) supersymmetric, whereas
	\eqref{sigma_phys} does break supersymmetry.
	It is imperative to find a way to introduce one extra mass parameter in the geometric formulation.

\subsection{Spontaneous Supersymmetry-breaking Geometric \\
Formulation}
\label{sbreaking}

	Since $ \mc{B} $ is a twisted superfield, one can introduce a twisted superpotential of the form
\[
	\frac{\Delta\mc{L}^\text{tw.m.}_{(0,2)}}{2\beta} ~~\supset~~
	\frac{i}{\sqrt{2}}\,a \int \mc{B}\, d^2\wt{\theta} ~~+~~ \text{h.c.},
	\qquad\qquad\qquad\qquad d^2\wt{\theta} ~=~ d\ov{\theta}{}_L\,d\theta_R\,
\]
	(here $ i/\sqrt{2} $ is a convenient normalization factor).
	This creates a linear in $ \ff $ contribution in the Lagrangian
\[
	\frac{\Delta\mc{L}_{(0,2)}^\text{tw.m.}}
             {2\beta} ~~\supset~~ i\,a\,\ff ~+~ i\,\ov{a}\,\bff ~+~ \bff\,\ff ~+ \dots,
\]
	and, correspondingly, changes the $ \ff $-term condition \eqref{ffterm} to the following:
\[
	\ff ~~=~~ -\, \btgamma\, g_{i\bj}\, \psi_L^i\, \bpsi_R^\bj 
		~+~ i\, \btgamma\, \ov{m}_G^i \frac{\bphi^i\, \phi^i}{\chi}
		~-~ i\, \ov{a}\,.
\]
	Substituting this into Eq.~\eqref{geomlin} produces (a) vacuum energy, and (b) mass shifts both for
	bosons and fermions.
	Choosing the appropriate value 
$$ a = -\, \tgamma\,m^N \,,$$ 
	one can now match the masses of elementary excitations  to those of the gauge formulation.
	Overall, the supersymmetry-breaking theory has the following mass terms:
\begin{align}
\notag
	\frac{\mc{L}_{(0,2)}^\text{tw.m.}}{2\beta} & 
	~~=~~ \int\, d^4\theta\, \lgr K(\Phi,\ov{\Phi}, V^i) 
		~-~ 2\, \ov{\mc{B}}\,\mc{B}  
		~-~  \sqrt{2}\, \tgamma\,\mc{B}\,K  ~-~ \sqrt{2}\, \ov{\tgamma}\,\ov{\mc{B}}\,\ov{K} \rgr
	\\[0.5mm]
\notag
	&\!\!\!\!\!\!\!\!
	~-~ \frac{i}{\sqrt{2}} \int d^2\wt{\theta} \cdot \tgamma\,m^N\,\mc{B} 
	~-~ \frac{i}{\sqrt{2}} \int d^2\ov{\wt{\theta}} \cdot \btgamma\, \ov{m}{}^N\, \ov{\mc{B}}
	~~\supset~~
	\\
\notag
	&\!\!\!\!\!\!\!\!
	~\supset~~
	|\tgamma|^2\, \left|m^N\right|^2 
	~+~ g_{i\bj}\, m_{G\mu}^i\, m_{G\mu}^\bj\, \phi^i\, \bphi^\bj
	~+~ |\tgamma|^2\, \lgr \ov{m}{}^N m_G^i ~+~ m^N \ov{m}{}_G^i \rgr \frac {\phi^i\,\bphi^i}{\chi}
	\\[1mm]
\label{geom_mass}
	&\!\!\!\!\!\!\!\!
	~+~ |\tgamma|^2 \cdot \ov{m}{}_G^\bj\,\frac{\bphi^\bj\,\phi^\bj}{\chi}
			\cdot m_G^i\, \frac{\bphi^i\, \phi^i}{\chi}
	\\[1mm]
\notag
	&\!\!\!\!\!\!\!\!
	~+~ i\,\frac{1}{2}\,g_{i\bj}\,\psi_L^i\, \overleftrightarrow{m}{}_{\!G} \bpsi_R^\bj
	~+~ i\,\frac{1}{2}\,g_{i\bj}\,\psi_R^i\, \overleftrightarrow{\ov{m}}{}_{\!G} \bpsi_L^\bj
	\\[3mm]
\notag
	&\!\!\!\!\!\!\!\!
	~+~ i\,|\tgamma|^2\,m^N\,g_{i\bj}\,\psi_L^i\, \bpsi_R^\bj
	~+~ i\,|\tgamma|^2\,\ov{m}{}^N\,g_{i\bj}\,\psi_R^i\, \bpsi_L^\bj
	\\[4mm]
\notag
	&\!\!\!\!\!\!\!\!
	~+~ i\tgamma\, g_{i\bj}\, m_G^\bj\,\bphi^\bj\, \psi_L^i\,\zr
	~-~ i\btgamma\, g_{i\bj}\,\ov{m}{}_G^i\, \bpsi_L^\bj\, \phi^i\,\bzr
	\\[1mm]
\notag
	&\!\!\!\!\!\!\!\!
	~+~ i\, |\tgamma|^2\, (g_{i\bj}\,\psi_L^i\,\bpsi_R^\bj)\, \cdot m_G^k \frac{\bphi^k\, \phi^k}{\chi}
	~+~ i\, |\tgamma|^2\, (g_{i\bj}\,\psi_R^i\,\bpsi_L^\bj)\, \cdot \ov{m}{}_G^\bk \frac{\bphi^\bk\,\phi^\bk}{\chi}
	\,.
\end{align}
With all $N$ mass parameters included, the (spontaneous) breaking of SUSY occurs right away, 
at the classical level.

Under the stereographic projection \eqref{stereo} this turns into
\begin{align*}
	\frac{\mc{L}_{(0,2)}^\text{tw.m.}}{2\beta} & 
	~~\supset~~ 
	|\tgamma|^2\, \left|m^N\right|^2 
	\\
	&\!\!\!\!\!\!\!\!
	~+~
	\lgr |m_G^i|^2 ~+~
		|\tgamma|^2 \Bigl\{ \ov{m}{}^N m_G^i ~+~ m^N \ov{m}{}_G^i \Bigr\} \rgr \left|n^i \right|^2 
	~-~ (1 \!-\! |\tgamma|^2)\, \Bigl| \sum_i m_G^i\, |n^i|^2 \Bigr|^2 
	\\
	&\!\!\!\!\!\!\!\!
	~-~ i\, ( m_G^i \,+\, |\tgamma|^2 m^N )\, \bxi_{Ri}\, \xi_L^i
	~-~ i\, (\ov{m}{}_G^i \,+\, |\tgamma|^2 \ov{m}{}^N )\, \bxi_{Li}\, \xi_R^i
	\\[4mm]
	&\!\!\!\!\!\!\!\!
	~+~ i\,(1\!-\!|\tgamma|^2)\, m_G^i\,|n^i|^2\, (\bxir\,\xil) 
	~+~ i\,(1\!-\!|\tgamma|^2)\, \ov{m}{}_G^i\,|n^i|^2\, (\bxil\,\xir)
	\\[4mm]
	&\!\!\!\!\!\!\!\!
	~+~ i\,\tgamma\, m_G^i\,\ov{n}{}_i\, \xi_L^i\, \zr
	~-~ i\,\btgamma\, \ov{m}{}_G^i\, \bxi_{Li}\, n^i\, \bzr\,,
	\\[5mm]
	&\!\!\!\!\!\!\!\!
	~-~ i |\tgamma|^2\, m^N\, \ov{\xi}{}_{RN}\, \xi_L^N
	~-~ i |\tgamma|^2\, \ov{m}^N\, \ov{\xi}{}_{LN}\, \xi_R^N\,,
	\qquad\qquad
	i ~=~ 1,...\,N-1\,.
\end{align*}
	We observe that this Lagrangian matches exactly
	onto the gauge formulation of the heterotic massive sigma model \eqref{sigma_mass}, provided that
	we accept
\[
	m_G^i ~~=~~ m^i ~-~ m^N \,.
\]
As an additional check we now show that in the 
large  mass limit, $|m^l|\gg \Lambda$, we can
recover all $N$ Higgs vacua (\ref{higgsn}) obtained in the
gauge formulation from the geometric formulation (\ref{geom_mass}). One of these vacua (with $l_0=N$) corresponds
to $\phi^i=0$. Other $(N-1)$ vacua are located at $\phi^{l_0}\to\infty$, $l_0=1,...,(N-1)$ as  seen from 
(\ref{stereo}). The vacuum energies and boson and fermion masses 
in these vacua exactly match expressions (\ref{hetmass})
obtained in the gauge formulation.

	While it took us some effort to prove \ntwoo supersymmetry of the theory \eqref{sigma_full},
	the geometric formulation of this theory 
\begin{align*}
	\frac{\mc{L}_{(0,2) }^\text{tw.m.}}{2\beta} & 
	~~=~~ \int\, d^4\theta\, \lgr K(\Phi,\ov{\Phi}, V^i) 
		~-~ 2\, \ov{\mc{B}}\,\mc{B}  
		~-~  \sqrt{2}\, \tgamma\,\mc{B}\,K  ~-~ \sqrt{2}\, \ov{\tgamma}\,\ov{\mc{B}}\,\ov{K} \rgr
	\\
	&~
	~-~ \frac{i}{\sqrt{2}} \int d^2\wt{\theta} \cdot \tgamma\,m^N\,\mc{B} 
	~-~ \frac{i}{\sqrt{2}} \int d^2\ov{\wt{\theta}} \cdot \btgamma\, \ov{m}{}^N\, \ov{\mc{B}}
\end{align*}
	is manifestly supersymmetric.

\section{Conclusions}
\label{con}

In this paper we considered various two-dimensional supersymmetric sigma models on the CP$(N-1)$
target space. We constructed an \ntwoo  model which combines the twisted mass deformation 
with the heterotic deformation following from
the bulk theory deformation (\ref{defpo}). Rather unexpectedly, in addition to the heterotic coupling parameter,
this model  contains an``extra" mass parameter which was unobservable in the absence of the heterotic deformation.
If all $N$ mass parameters are nonvanishing, \ntwoo supersymmetry is (spontaneously) broken at
the tree level. Setting one  mass parameters to zero, we restore supersymmetry at the classical 
level. At the quantum (nonperturbative) level it is still spontaneously broken.

There are two obvious tasks for the nearest future. First, the \ntwoo  model   combining the twisted mass deformation 
with the heterotic deformation must be fully derived from the microscopic \none Yang--Mills theory in the bulk.
Second, it must be solved in the large-$N$ limit. The solution of both problems is withing reach.

\section*{Acknowledgments}
The work of PAB was supported in part by the NSF Grant No. PHY-0554660. PAB is grateful for kind
hospitality to FTPI, University of Minnesota, where part of this work was done. 
The work of MS is supported in part by DOE grant DE-FG02-94ER408. 
The work of AY was  supported 
by  FTPI, University of Minnesota, 
by RFBR Grant No. 09-02-00457a 
and by Russian State Grant for 
Scientific Schools RSGSS-11242003.2.

\newpage

\addcontentsline{toc}{section}{Appendices}
\setcounter{section}{0}
\renewcommand{\thesection}{\Alph{section}}

\section{Notations in Euclidean Space}
\label{app:eucl}
%
%

\setcounter{equation}{0}

Since \cpn sigma model can be obtained as a dimensional reduction from four-dimensional theory,
we present first our four-dimensional notations.
The indices of four-dimensional spinors are raised and lowered by the SU(2) metric tensor,
\beq
	\psi_\alpha ~=~ \epsilon_{\alpha\beta}\, \psi^\beta, \qquad
	\ov{\psi}{}_{\dot{\alpha}} ~=~ \epsilon_{\dot{\alpha}\dot{\beta}}\, \ov{\psi}{}^{\dot\beta}, \qquad 
	\psi^\alpha ~=~ \epsilon^{\alpha\beta}\, \psi_\beta, \qquad
	\ov{\psi}{}^{\dot{\alpha}} ~=~ \epsilon^{\dot\alpha\dot\beta}\, \ov{\psi}{}_{\dot\beta}~,
\eeq
	where
\beq
	\epsilon_{\alpha\beta} ~=~ \epsilon_{\dot\alpha\dot\beta} ~=~
			\lgr \begin{matrix}
			     	\ 0\  &  \ 1\   \\
				 -1\  &  \ 0\  
			     \end{matrix} \rgr,
	\qquad \text{and} \qquad
	\epsilon^{\alpha\beta} ~=~ \epsilon^{\dot\alpha\dot\beta} ~=~
			\lgr \begin{matrix}
				\ 0\ &   -1\   \\
				\ 1\ &  \ 0\ 
			     \end{matrix} \rgr.
\eeq
The contractions of the spinor indices are short-handed as
\beq
	\lambda\psi ~=~ \lambda_\alpha\, \psi^\alpha\,, \qquad
	\ov{\lambda\psi} ~=~  \ov{\lambda}{}^{\dot\alpha}\, \ov{\psi}{}_{\dot\alpha}\,.
\eeq
The sigma matrices for the euclidean space we take as
\beq
	\sigma^{\alpha\dot\alpha}_\mu ~=~  \lgr {\bf 1},\quad -i\,\tau^k \rgr^{\alpha\dot{\alpha}}
	\hspace{-2.0ex},
	\qquad
	\ov{\sigma}{}_{\dot\alpha\alpha\, \mu} ~=~ 
			\lgr {\bf 1},\quad i\, \tau^k \rgr_{\dot\alpha\alpha},
\eeq
where $ \tau^k $ are the Pauli matrices.

Reduction to two dimensions can be conveniently done by picking out $ x^0 $ and $ x^3 $ 
as the world sheet (or ``longitudinal'') coordinates, and integrating over the orthogonal coordinates. 
The two-dimensional derivatives are the defined to be
\beq
	\p_R  ~~=~~ \p_0 ~+~ i\p_3\,, \qquad   \p_L ~~=~~ \p_0 ~-~ i\p_3\,.
\eeq
One then identifies the lower-index spinors as the two-dimensional left- and right-handed chiral spinors
\beq
	\xi_{R} ~=~ \xi_{1}\,, \quad\qquad
	\xi_{L} ~=~ \xi_{2}\,, \quad\qquad\qquad
	\ov{\xi}{}_{R} ~=~ \ov{\xi}{}_{\dot{1}}\,, \quad\qquad
	\ov{\xi}{}_{L} ~=~ \ov{\xi}{}_{\dot{2}}\,.
\eeq

For two-dimensional variables, the CP($N-1$) indices are written as upper ones 
\[
	n^l\,, \quad \xi^l\,,
\]
and lower ones for the conjugate moduli
\[
	\ov{n}{}_l\,, \quad \ov{\xi}{}_l\,, 
\]
where $ l~=~1,\, ...,\, N $.
In the geometric formulation of CP($N-1$), global indices are written upstairs in both cases, only
for the conjugate variables the indices with bars are used 
\[
	\phi^i\,,\ \psi^i\,, \qquad \ov{\phi}{}^\bi\,,\ \ov{\psi}{}^\bi\,, 
	\qquad\qquad i,\ \bi ~=~ 1,\,...,\,N-1\,,
\]
and the metric $ g_{i\bj} $ is used to contract them.


\section{Minkowski versus Euclidean formulation}
 \renewcommand{\theequation}{\Alph{section}.\arabic{equation}}
\setcounter{equation}{0}
 
 \renewcommand{\thesubsection}{\Alph{section}.\arabic{subsection}}
\setcounter{subsection}{0}
\label{app:mink}

Although this work considers the formulation of heterotic \cpn model in Euclidean space only, 
a series of our papers work with both Minkowski and Euclidean conventions \cite{SYhet,BSYhet,SYhetlN}.
It is useful to summarize the transition rules.
If the Minkowski coordinates are
\beq
x^\mu_M =\{t,\,z\}\,,
\label{appeone}
\eeq
the passage to the Euclidean space requires
\beq
t \to - i\tau\,,
\label{appe2}
\eeq
and the Euclidean coordinates are
\beq
x^\mu_M =\{\tau,\,z\}\,.
\label{appe3}
\eeq
The derivatives are defined as follows:
\beqn
\pt_L^M &=& \pt_t+\pt_z\,,\qquad \pt_R^M = \pt_t- \pt_z\,,
\nonumber\\[2mm]
\pt_L^E &=& \pt_\tau - i \pt_z\,,\qquad \pt_R^E = \pt_\tau + i \pt_z\,.
\label{appe4}
\eeqn
The Dirac spinor is
\beq
\Psi =\left(
\begin{array}{c}
\psi_R\\[1mm]
\psi_L
\end{array}
\right)
\label{appe5}
\eeq
In passing to the Eucildean space $\Psi^M = \Psi^E$;
however, $\bar\Psi$ is transformed,
\beq
\bar\Psi^M \to i \bar \Psi^E\,.
\label{appe6}
\eeq
Moreover, $\Psi^E$ and $\bar \Psi^E$ are {\em not} related by the complex conjugation operation.
They become independent variables. The fermion gamma matrices are defined as
\beq
\bar\sigma^\mu_M =\{1,\,-\sigma_3\}\,,\qquad \bar\sigma^\mu_E =\{1,\, i\sigma_3\}\,.
\label{appe7}
\eeq
Finally, 
\beq
\cell_E =- \cell_M (t=-i\tau , ...).
\eeq
With this notation, formally, the fermion kinetic terms in $\cell_E $ and $\cell_M $
coincide.
We use the following equivalent definitions of the heterotic deformation terms
\beq
 \frac{g_0}{\sqrt 2} \tilde{\gamma}_\ssm\, \zeta_R G_{i\bj}\left(i\pt_L\bar\phi^{\,\bj}\right)\psi_R^i\,, 
\qquad
  \frac{1}{g_0^2}\, \wt{\gamma}_\sse\, \chi_R^a\, (i\p_L S^a) \zeta_R\,,
\qquad
  \frac{2}{g_0^2}\, \wt{\gamma}_\sse\, (i\p_L \ov{n}) \xi_R \zeta_R\,
\label{appabiferm}
\eeq
in Minkowski and Euclidean spaces correspondingly. 
The following transition rule applies,
\beq
\wt{\gamma}_\ssm = - i\,\wt\gamma_\sse\,.
\label{appe8}
\eeq
Everywhere where there is no menace of confusion we omit the super/sub\-scripts $M,E$.
The first two terms in Eq.~(\ref{appabiferm}) originally were introduced in Ref.~\cite{SYhet}, with 
a constant
\beq
	\gamma ~=~ \wt\gamma / (\sqrt{2}g_0) \,.
\eeq
In this paper, subscript $\scriptstyle (E)$ is always assumed for $\wt\gamma$.


\end{document}